# Consumer Law for AI Agents

*Christoph Busch*[*]

Osnabrück University, European Legal Studies Institute, Osnabrück, Germany
Email: christoph.busch@uos.de

**Abstract**

Since the public release of ChatGPT in November 2022, the AI landscape is undergoing a rapid transformation. Currently, the use of AI chatbots by consumers has largely been limited to image generation or question-answering language models. The next generation of AI systems, AI agents that can plan and execute complex tasks with only limited human involvement, will be capable of a much broader range of actions. In particular, consumers could soon be able to delegate purchasing decisions to AI agents acting as "Custobots". Against this background, the Article explores whether EU consumer law, as it currently stands, is ready for the rise of the Custobot Economy. In doing so, the Article makes three contributions. First, it outlines how the advent of AI agents could change the existing e-commerce landscape. Second, it explains how AI agents challenge the premises of a human-centric consumer law which is based on the assumption that consumption decisions are made by humans. Third, the Article presents some initial considerations how a future consumer law could look like that works for both humans and machines.

---

[*] This research has been carried out within the scope of the project "Granular Society – Granular Law? Individuality and Normative Models in the Data Society" funded by VolkswagenStiftung under a Momentum Grant. For valuable input, I am grateful to participants at the 2024 Regulating AI Conference at Peking University of Transnational School of Law, the 2025 Consumer Law Scholars Conference at Boston University as well as workshops hosted at the University of Münster, Keio University, Tokyo, and Korea University, Seoul.

## A. Introduction

On January 23, 2025, OpenAI, the company behind ChatGPT released a preview version of "Operator", an AI agent that can browse the web and perform tasks for users.[1] Operator can interact with websites and automate a variety of actions like buying goods online, booking train tickets and making restaurant reservations. Unlike AI chatbots, agentic AI systems such as Operator are not limited to image generation or answering questions. They can interact with digital environments and take actions on behalf of users. In particular, consumers could soon be able to delegate their purchasing decisions to AI agents. Sooner rather than later, AI agents will interact autonomously with apps or online shops and purchase goods and services in accordance with pre-defined preferences and payment options.

Against this background, this Article argues that EU consumer law needs to prepare for the rise of AI agents such as OpenAI's Operator and even more powerful AI tools that might become available in the near future. Consumer law, as it currently stands, is human-centric. It is based on the premise that decisions to buy goods or services are made by humans. In the future, however, a growing number of consumption decisions will be made by AI agents acting as "algorithmic consumers"[2], "machine customers"[3] or "Custobots"[4] (a portmanteau of 'customer' and 'robot'). Consumer policy makers need a plan to update the existing regulatory framework to enable consumers to capture the benefits of AI agents and to mitigate potential risks arising in the context of agentic commerce.

---

[1] Introducing Operator, OpenAI (Jan. 23, 2025), https://openai.com/index/introducing-operator/

This impending shift in consumer decision-making from humans to AI agents requires several adjustments to the regulatory design of consumer law. Currently, consumer law is geared towards enabling informed decision-making by human consumers. In the future, however, the specific strengths and weaknesses of non-human consumers must be taken into account. As a consequence, for example, the design of pre-contractual disclosures has to be reconsidered as AI agents can process larger amounts of information. Similarly, the rules on unfair commercial practices need to be recalibrated as the vulnerability of AI agents to misleading and aggressive commercial practices may differ from that of humans. On the one hand, AI agents will be less vulnerable to dark patterns and emotional advertising messages. On the other hand, new AI-specific vulnerabilities could arise such as prompt injections and other adversarial attacks that trick AI agents into a transaction.

This article adds to the expanding literature on AI agents. Other scholars have analyzed legal issues regarding AI agents through the lens of agency law[5] or they focus primarily on issues of liability, antitrust or macro-level effects of AI agents.[6] This article chooses a different perspective. Building on the work of the European Law Institute's project on Algorithmic Contracts,[7] this Article explores whether EU consumer law, as it currently stands, is ready for the rise of AI agents. In doing so, the Article seeks to identify implications for consumer policy, and motivate further empirical research.

The remainder of the Article proceeds as follows. Part B provides a brief overview of the recent development from generative AI to agentic AI and how AI agents will change the functioning of consumer markets. Part C turns to the question what the rise of AI agents means for consumer law. In particular, it will be considered how the emergence of the Custobot Economy will challenge the premises of consumer law. Part D seeks to identify core elements that could contribute to a consumer law that works for both humans and machines. Part E concludes.

---

[2] Michal S. Gal & Niva Elkin Koren, Algorithmic Consumers, 30 Harv. J. L. & Tech. 309 (2017).
[3] Don Scheibenreif & Mark Raskino, When Machines Become Customers, 2023.
[4] Scheibenreif & Raskino, *supra* note 3, at 1; see also Christian Twigg-Flesner & Geraint Howells, Adapting Consumer Law to New Technologies, in Roger Brownsword & Larry DiMatteo (eds) The Cambridge Handbook of the Governance of Technology: Discontent, Disconnect and Disruption (forthcoming 2025).
[5] See e.g. Noam Kolt, Governing AI Agents (Feb. 11, 2025), 101 Notre Dame Law Review (forthcoming), http://dx.doi.org/10.2139/ssrn.4772956 (exploring the implications of agency law and theory for designing and regulating AI agents); see also Deven R. Desai & Mark Riedl, Responsible AI Agents (Feb. 20, 2025), http://dx.doi.org/10.2139/ssrn.5147666 (arguing that most of the concerns around AI agents should be framed as what computer science broadly calls the „alignment problem").
[6] See, e.g. Gal & Elkin-Koren, *supra* note 2; Rory van Loo, Digital Market Perfection, 117 Michigan Law Review 815, 830 (2019).
[7] European Law Institute, EU Consumer Law and Automated Decision-Making (ADM): Is EU Consumer Law Ready for ADM? (December 2023), Interim Report of the European Law Institute, https://www.europeanlawinstitute.eu/fileadmin/user_upload/p_eli/Publications/ELI_Interim_Report_on_EU_Consumer_Law_and_Automated_Decision-Making.pdf. The author served as co-reporter for the ELI project on Algorithmic Contracts.

## B. The Custobot Economy

This Part outlines the key technological forces driving the rise of the so-called "Custobot Economy," which will differ in important ways from today's e-commerce landscape. It focuses in particular on the emergence of AI agents, explaining how these systems differ from earlier generative AI models like ChatGPT. It also explores how this shift toward more autonomous, agent-driven technologies – the "agentic turn" in electronic commerce – could transform the future of online shopping.

### I. From Chatbots to Custobots

Since the public release of ChatGPT in November 2022, generative AI has taken the world by storm. ChatGPT, which reached more than 100 million monthly users in just two months, is widely seen as the fastest-growing consumer internet app of all time.[8] More recently, generative AI tools have become relevant for e-commerce. Consumers can ask tools like OpenAI's ChatGPT, Microsoft's Copilot, or Google's Gemini for advice, product recommendation or suggestions where to buy goods and services. In November 2024, the AI-driven search engine Perplexity introduced a new feature that allows Pro subscribers to purchase a product without leaving its AI search engine. Users can click a "Buy with Pro" button that will automatically order the product on the basis of saved shipping and billing information.[9]

AI agents like OpenAI's Operator are taking the shopping integration one step further. In the words of Jonathan Zittrain, the advent of AI agents marks "a crossing of the blood-brain barrier between digital and analog, bits and atoms".[10] Unlike AI-driven chatbots that are designed to *assist* users in their transactional decisions, AI agents can *act* on behalf of users. Some AI agents need specific application programming interfaces (API) to interact with other systems,[11] others can autonomously navigate websites like humans. While a chatbot is merely suggesting things a consumer could do, an AI agent can do the things it suggests for the user. Instead of making suggestions how to travel to the airport, the AI agent selects a suitable travel option and orders, for example, an Uber. From a consumer perspective the evolution from generative AI to agentic AI could be described as a transition from "augmented consumers" to "algorithmic consumers".[12]

Before exploring how AI agents will change the way e-commerce works, we first need to clarify what the term AI agent means. There is not yet a generally agreed and definitive definition of AI agent.[13] From a technology perspective, AI agents are to be distinguished from generative AI models, such as Large Language Models (LLMs) or other foundation models (FMs), although AI agents make use of such AI models to perform their tasks.[14]

---

[8] Catherine Thorbecke, A year after ChatGPT's release, the AI revolution is just beginning (CNN Business, November 30, 2023), https://edition.cnn.com/2023/11/30/tech/chatgpt-openai-revolution-one-year/index.html.
[9] Emma Roth, Perplexity's AI search engine can now buy products for you (Nov. 18, 2024), https://www.theverge.com/2024/11/18/24299574/perplexity-ai-search-engine-buy-products.
[10] Jonathan Zittrain, We Need to Control AI Agents Now, The Atlantic (Jul. 2, 2024), https://www.theatlantic.com/technology/archive/2024/07/ai-agents-safety-risks/
[11] See Desai & Riedl, *supra* note 5, at 27, ("What makes an LLM "agentic" is its ability to call APIs to other sytems.").
[12] Gal & Elkin-Koren, *supra* note 2.
[13] For an overview see Melissa Heikkilä, What Are AI Agents?, MIT Tech. Rev. (Jul. 5, 2024), https://www.technologyreview.com/2024/07/05/1094711/what-are-ai-agents/.
[14] Friso Bostoen & Jan Krämer, AI agents and ecosystem contestability, CERRE Issue paper (Nov. 2024), 5.

For the purposes of the following analysis, a working definition borrowed from a recent report on AI regulation by the UK Department for Science, Innovation and Technology may provide a conceptual basis: AI agents are "autonomous AI systems that perform multiple sequential steps—sometimes including actions like browsing the internet, sending emails, or sending instructions to physical equipment—to try and complete a high-level task or goal."[15] Similar definitions for AI agents (sometimes also referred to as 'agentic systems' or 'copilots') have been proposed by industry stakeholders, such as IBM[16] ("a system or program that is capable of autonomously performing tasks on behalf of a user or another system by designing its workflow and utilizing available tools"), or Open AI[17] ("AI systems that can pursue complex goals with limited direct supervision").[18] AI agents being used for purchasing decisions can be referred to as "machine customers" or "Custobots".

## II. The Rise of Agentic Commerce

The rise of AI agents represents a rare moment of technological disruption—one with the potential to fundamentally reshape how consumers engage with online marketplaces. This shift towards agent-driven commerce may prompt a sweeping reconfiguration of the current e-commerce landscape. In turn, it raises key conceptual questions about the nature and mechanics of online shopping, suggesting that some foundational assumptions about how digital markets function may need to be reconsidered.

To be sure, e-commerce has evolved continually since Amazon and eBay first launched their online marketplaces in 1995. Over time, user interface design has undergone several shifts, moving from early websites to mobile commerce, social commerce, and more recently, live shopping on platforms like TikTok. Yet throughout these waves of technological change and experimentation with business models, one element has remained constant: user interfaces were always designed to persuade *human* consumers to make purchasing decisions.

This core feature of e-commerce is likely to change with the advent of AI agents and the rise of agentic commerce ("a-commerce").[19] As human consumers increasingly delegate purchasing decisions to autonomous AI systems, the focus of online marketplaces will move from persuading humans to influencing AI agents. This transition is set to introduce several fundamental changes to the structure of e-commerce—each carrying significant implications from a consumer policy perspective.

1. **New Customer Journeys**

First of all, AI agents will change the way customers search for products. Today, the customer journey often begins with a search engine like Google or one of the large online marketplaces, such as Amazon, which has itself become a kind of product search engine. In the near future, AI agents could serve as new access points for commerce.[20] Deutsche Telekom recently presented the concept of an "app-less" AI phone which gives an idea of how this new generation of user interfaces might look like.[21] On the so-called "T-phone", users interact with service providers through an AI agent rather than individual apps. Users present their questions to the AI agent, which autonomously selects the underlying apps or service providers needed to fulfill the query. Rather than starting their customer journey at a specific website, app or search engine consumers will instruct their AI agent to find a product that matches their preferences and order it on their behalf. In this sense, AI agents are not merely a new "technology".[22] They will change how consumers interact with websites and online shops. From a competition perspective, this change has the potential to disrupt the "gatekeepers" that are controlling today's access points to online retail.[23]

---

[15] UK Department for Science, Innovation and Technology, A pro-innovation approach to AI regulation: government response (Feb. 6, 2024), https://www.gov.uk/government/consultations/ai-regulation-a-pro-innovation-approach-policy-proposals/outcome/a-pro-innovation-approach-to-ai-regulation-government-response; see also Bostoen & Jan Krämer, *supra* note 14, at 5.
[16] Anna Gutowska, What are AI agents?, IBM (Jul. 3, 2024), https://www.ibm.com/think/topics/ai-agents
[17] OpenAI, Practices for Governing Agentic AI Systems (Dec. 14, 2023) https://openai.com/index/practices-for-governing-agentic-ai-systems/.
[18] Bostoen & Jan Krämer, *supra* note 14, at 5.
[19] David G. W. Birch, A-Commerce Is Coming: Agentic AI And The "Do It For Me" Economy, Forbes (Feb. 6, 2025), https://www.forbes.com/sites/davidbirch/2025/02/06/a-commerce-is-coming-agentic-ai-and-the-do-it-for-me-economy/.

The developments outlined here could have far-reaching consequences for the design of online shops. The internet, website and apps largely have been built for use by humans and optimized to improve user experience for human customers. However, the shift towards agent-to-website interaction or agent-to-agent interaction could require major changes in the design of user interfaces. In particular, the content offered via websites and apps will need to focus more on information that AI agents can use to perform their tasks. In this sense, customer-centricity will require to understand how AI agents interact with websites and apps.

It remains to be seen what specific impact this will have on the design of online shops. One possible scenario is that that websites and apps will become the target for both human and agentic traffic. It is also conceivable that agent-to-agent exchanges could be driven by API connections which will be more scalable and can be better adapted to the information needs of AI agents. As a consequence, a new category of online shops for Custobots could emerge that resemble "dark supermarkets" or "ghost kitchens" specifically designed to cater for delivery services.

## 2. Comparison Shopping on Steroids

Today, most consumers shop a narrow set of brands and retailers.[24] One possible reason for consumer loyalty may be the widespread status quo bias exhibited by consumers.[25] Another reason may be that the search costs for extensive comparison shopping are simply too high. Therefore, it may be rational to limit the search radius for certain products. Yet another explanation could be that consumers seek to avoid choice overload as too many choices lead to less satisfaction with the decision and tend to lower decision quality.[26]

---

[20] Bostoen & Krämer, *supra* note 14, at 4.
[21] Deutsche Telekom showcases app-less AI smartphone concept, Reuters (Feb. 26, 2024), https://www.reuters.com/technology/deutsche-telekom-showcases-app-less-ai-smartphone-concept-2024-02-26/
[22] Cf. Jack M. Balkin, The Path of Robotics Law, 6 Cal Law Rev 45, 49 (2015) (arguing that "what we call the effects of technology are not so much features of things as they are features of social relations that employ those things").
[23] Bostoen & Krämer, *supra* note 14, at 4.
[24] Jur Gaarlandt, Wesley Korver & Andrew Shipilov, AI Agents Are Changing How People Shop. Here's What That Means for Brands, Harvard Business Review (Feb. 26, 2025), https://hbr.org/2025/02/ai-agents-are-changing-how-people-shop-heres-what-that-means-for-brands.
[25] See generally William Samuelson & Richard Zeckhauser, Status quo bias in decision making, J Risk Uncertainty 1, 7–59 (1988).
[26] See generally Barry Schwartz, The paradox of choice: Why more is less (HarperCollins Publishers 2004).

These problems could be mitigated through AI agents. While it is simply too overwhelming for human consumers to search everywhere for suitable products, AI agents can do this.[27] In addition, Custobots will be able to draw on a broader range of data (e.g. customer reviews, user forums, product tests, online media) and consider more factors when comparing products.[28] This could help consumers to reduce existing information asymmetries. Currently, some companies design their user interfaces in ways which (intentionally) complicate comparisons between different products. For example, Amazon does not allow users to sort products by unit price, requiring consumers to look through a large number of items to find the best price per unit.[29] In the near future, consumers can delegate this laborious task to their AI agent.

The AI-driven facilitation of comparison shopping will be particularly relevant for commodity items such as light bulbs or vacuum cleaner bags where brand visibility is limited or nonexistent during use.[30] Most likely, consumers will delegate the purchase of these categories of products to AI agents in the future. AI agents drawing on customer reviews and other relevant product data could help consumers not only to find the best offer for their preferred brand, they could also find lower-priced alternative products. While this could save consumers considerable time and money, the rise of AI agents will make it more difficult for brands to maintain customer loyalty against cheaper, often equally effective alternatives.[31]

## 3. Optimizing for AI Agents

How will brands and retailers react to these structural changes in the e-commerce landscape? Currently, retailers and manufacturers are trying to get as high up as possible in the rankings of search engines and online marketplaces. Empirical research suggests that "consumers are more likely to select options near the top of a list of results, simply by virtue of their position and independent of relevance, price or quality of the options"[32] As a consequence, appearing in the first position of a ranking offers online retailers a significant advantage. Similarly, click-through rates of ads on search engines decrease with positions.[33] Because of this „ranking effect or „position bias" brands and retails invest in search engine optimization (SEO) and ranking optimization as important tools for increasing sales.

In the new world of agentic commerce, this could change. If the purchasing decisions of AI agents are less influenced by ranking effects and position bias but rather based on an objective consideration of different factors, it could make less sense for brands and retailers to invest in SEO or to pay for a better ranking position on an online platform. Instead, it may be necessary to adopt different strategies. Rather than optimizing their visibility for search engines, businesses could try to optimize their marketing strategy for agentic consumption and increase their visibility for AI agents. As a consequence, current SEO practices could be replaced by a new strategy of of AI agent optimization (AAO).[34]

---

[27] See Van Loo, *supra* note 6, at 835 („A diligent consumer might look at five products on average before making a purchase, but AIs could look at thousands.").
[28] Gaarlandt, Korver & Shipilov, *supra* note 24.
[29] Rory van Loo, Helping Buyers Beware: The Need for Supervision of Big Retail, 163 U. Pa. L. Rev. 1311, 1345-47 (2015); see also Glenn Ellison & Sara Fisher Ellison, Search Obfuscation, and Price Elasticities on the Internet, 77 Econometrica 427, 428-29 (2009) (showing that in highly commodified electronic parts markets, consumers paid 6-9% higher prices because sellers were able to make product comparison more difficult).
[30] Gaarlandt, Korver & Shipilov, *supra* note 24.
[31] Id.
[32] UK Competition and Markets Authority, Algorithms: How they can reduce competition and harm consumers (Mar. 25, 2022), https://www.gov.uk/find-digital-market-research/algorithms-how-they-can-reduce-competition-and-harm-consumers-2021-cma; Raluca Mihaela Ursu, The Power of Rankings: Quantifying the Effect of Rankings on Online Consumer Search and Purchase Decisions, 37(4) Marketing Science 530-552 (2018); see also Christoph Busch, From Algorithmic Transparency to Algorithmic Choice: European Perspectives on Recommender Systems and Platform Regulation, in Sergio Genovesi, Katharina Kaesling & Scott Robbins (eds), Recommender Systems: Legal and Ethical Issues (Springer 2023) 31-54.
[33] Ashish Agarwal, Kartik Hosanagar, & Michael D. Smith, Location, location, location: An analysis of profitability of position in online advertising markets, 48(6) Journal of Marketing Research 1057-1073 (2011).

## C. How AI Agents Challenge the Premises of Consumer Law

Building on the analysis of the upcoming changes of the e-commerce landscape, this Part turns to the question why the rise of AI agents challenges the basic premises of consumer law and how consumer law rules must adapt to the new reality of AI driven markets. As outlined above, the advent of AI agents has the potential to fundamentally reshape how consumers make purchasing decisions or other transactional decisions (e.g. managing subscriptions). If built with the right legal and technical safeguards, AI agents could become a tool for consumer empowerment shifting the balance of power between businesses and consumers.[35] In so doing, AI agents could save consumers considerable time and money.[36] More specifically, AI agents could help consumers make better informed choices, and less biased purchasing decisions. Does this mean that AI agents will finally embody the theoretical model of the rational *homo economicus*? Not necessarily. The answer is more complicated, as will be explained in the following sections. In doing so, the focus will be on four topics: the average consumer concept, information duties, consumer choice and the concept of digital vulnerability.

### I. Beyond human-centricity: Towards an Average Consumer Test for Custobots?

For assessing the profound impact of AI agents on EU consumer law, it is necessary to consider first a fundamental premise of consumer law that seems so self-evident that it is usually not even stated. EU consumer law in its current form is *human-centric*, i.e. it is designed with the needs and weaknesses of *human* consumers in mind. Put differently, the design of consumer law rules is based on the premise that purchasing decisions are made by humans. Therefore, EU consumer law aims to protect human consumers and ensure their autonomy.

But this premise no longer holds in the Custobot Economy, which will be more complex and diverse. In the brave new world of AI-driven retail old-school human consumers will coexist with "augmented consumers" (who are assisted by AI tools such as ChatGPT) and AI agents acting as "algorithmic consumers".

The shift towards agentic AI systems could have a particular impact on the concept of the "average consumer". This consumer image or *Verbraucherleitbild* serves as a normative standard for the application of consumer law rules.[37] For example, an assessment of whether a commercial practice is unfair involves an examination of whether it materially distorts or is likely to materially distort the economic behavior of an average consumer.[38] To answer this question, the Court of Justice of the European Union (CJEU), takes as a benchmark the average consumer who is "reasonably well-informed, reasonably observant and circumspect".[39] This average consumer concept, which was initially developed by the case law of the CJEU, has since been codified by the European legislator in the secondary legislation.[40]

---

[34] Gaarlandt, Korver & Shipilov, *supra* note 24.
[35] Dazza Greenwood, Empowering Consumers with Personal AI Agents: Legal Foundations and Design Considerations, Consumer Reports (Oct. 18, 2024), https://innovation.consumerreports.org/empowering-consumers-with-personal-ai-agents-legal-foundations-and-design-considerations/.
[36] Rory van Loo, Digital Market Perfection, 117 Michigan Law Review 815, 830 (2019).
[37] See e.g. Peter Cartwright, The consumer image within EU law, in Christian Twigg-Flesner (ed) Research Handbook on EU consumer and contract law (Edward Elgar Publishing 2016) 199-220.
[38] See Art. 5(2) Unfair Commercial Practices Directive 2005/29/EC.

The average consumer concept has long been criticized by behavioralists as an unrealistic assumption.[41] Indeed, having a realistic understanding of consumer behavior is critical for getting regulation right. Therefore, it is no surprise that the interest in behavioral insights among consumer law scholars has increased significantly in recent years.[42] Responding to the behavioralist criticism, the CJEU, most recently, has underlined that the average consumer test is not 'supposed to be merely a theoretical exercise' and that 'considerations that are more realistic must also be taken into account'.[43] More specifically, the Court held that "the fact that the concept of 'average consumer' must be understood by reference to a consumer who is 'reasonably observant and circumspect' does not exclude taking into account the influence of cognitive biases on that average consumer, provided that it is established that such biases are likely to affect a reasonably well-informed and reasonably observant and circumspect person, to such an extent as to materially distort his or her behaviour."[44] Finally, it seems, the CJEU embraces a behavioralist approach to the average consumer test.

But how to apply such a behaviorally informed average consumer test to "algorithmic consumers"? How does the decision-making of AI agents differ from that of humans? Behavioral research shows that human consumers "tend to estimate poorly, have excess confidence in their decisions, and face difficulties processing even basic numerical decisions".[45] AI agents, in contrast, could be less vulnerable to such behavioral biases. For example, AI agents may be less subject to present bias than humans and will thus not suffer from hyperbolic discounting. Will this bring the end of teaser rates? Will AI agents help to solve problems related to buy-now-pay-later schemes and enable consumers to make better financial decisions? AI agents could also help consumers to overcome problems arising from status quo bias, overconfidence, and inattention. Unlike human consumers, AI agents will not forget to cancel a trial subscription, and they could remind a consumer to terminate the gym membership or the Xbox game pass that is no longer being used.[46] Considering these examples, one could assume that the behavior of AI agents might come closer to the elusive ideal of the reasonably well-informed, reasonably observant and circumspect consumer. From this perspective, the normative model which has often been criticized as an unrealistic assumption, could become reality with the advent of AI agents.

---

[39] CJEU, Case C-210/96 Gut Springenheide and Rudolf Tusky v Oberkreisdirektor des Kreises Steinfurt [1998] ECR I – 4657, paras 30-31.
[40] See Recital 18 Unfair Commercial Practices Directive 2005/29/EC; see also Hanna Schebesta & Kai P. Purnhagen, Island or Ocean: Empirical Evidence on the Average Consumer Concept in the UCPD, 28 European Review of Private Law 293-310 (2020).
[41] See, e.g. Rossella Incardona & Cristina Poncibò, The average consumer, the unfair commercial practices directive and the cognitive revolution, 30 Journal of Consumer Policy 21 (2007).
[42] For an overview see Fabrizio Esposito, Conceptual foundations for a European consumer law and behavioural sciences scholarship in Hans-W. Micklitz, Anne-Lise Sibony & Fabrizio Esposito (eds) Research Methods in Consumer Law (Edward Elgar Publishing 2018) 38-76.
[43] CJEU, Case C-646/22 Compass Banca SpA v Autorità Garante della Concorrenza e del Mercato (Nov. 14, 2024), para. 51.
[44] Id. at para. 53.
[45] Van Loo, *supra* note 6, at 833; see also Christine Jolls, Behavioral Economics Analysis of Redistributive Legal Rules, 51 Vand. L. Rev. 1653, 1659 (1998); Christine Jolls, Cass Sunstein and Richard Thaler, A Behavioral Approach to Law and Economics, 50 Stan. L. Rev. 1471.
[46] Cf. Ulrike Malmendier & Stefano DellaVigna, Paying Not to Go to the Gym, 96(3) American Economic Review 694 (2006) (showing that overconfident consumers overestimate gym attendance as well as the cancellation probability of automatically renewed contracts); see also Christoph Busch & Christian Twigg-Flesner, A Roadmap for Regulating Subscriptions in the Digital Fairness Act, Journal of European Consumer and Market Law 234 (2024) (providing policy suggestions on how to regulate digital subscriptions).

On closer inspection, however, things are more complicated. Recent research in the emerging fields of „machine psychology"[47] and "AI psychometrics",[48] which study the behavior of large language models (LLMs), such as those on which AI agents are based, indicates that some LLMs exhibit reasoning errors similar to those ascribed to heuristic-based human reasoning.[49] So far, however, the findings are somewhat inconclusive. In a recent study, LLMs showed human-like decision biases (e.g. risk aversion, confirmation bias) in one half of the experiments and appeared less biased in the other half.[50] In the literature, two possible sources for emergent cognitive biases of LLMs are being discussed: One possibility is that the models have learned the biases from the corpus of training data which is generated by bounded rational humans; another possibility could be that the biases are introduced through human feedback-based fine-tuning of the models.[51]

In summary, at this stage, it is an open question whether AI-driven Custobots will be more "rational" than human consumers. We currently still know too little about the behavior of AI agents to be able to give a definite answer as to whether the "average consumer test" developed for humans can also be applied to AI agents, or whether an "average Custobot test" needs to be developed.[52] One problem seems to be that AI agents do not have a stable and persistent "AI persona". Whether an AI agent is risk averse or risk taking very much depends on the respective system prompt and the specific user prompts.[53] Recent research even indicates that some AI agents exhibit a tendency to cheat in order to achieve their goal.[54] Maybe the brave new world of agentic commerce will not only need answers to rogue traders, but also AI-driven "rogue consumers".

## II. Information Duties: More is More?

The rise of agentic commerce is not only a challenge for the average consumer concept. Another pillar of consumer law that could be shaken by the rise of AI agents is the "information paradigm".[55] Information duties or mandated disclosures are one of the most widely used tools in consumer law. Long lists of standardized information duties are a signature feature of EU consumer law directives.[56] The underlying idea is that mandated disclosures will enable consumers to take informed and therefore better decisions. However, a growing body of behavioral research has questioned the effectiveness of mandated disclosures and argues that the large number of pre-contractual disclosures often lead to information overload making consumer decisions not better, but worse.[57] In addition to the quantitative problem (information overload), there is often also a qualitative problem: a large proportion of the standardized information provided is not relevant for the individual consumer (information mismatch).

---

[47] Thilo Hagendorff, Ishita Dasgupta, Marcel Binz, Stephanie C.Y. Chan, Andrew Lampinen, Jane X. Wang, Zeynep Akata, Eric Schulz, Machine Psychology, https://doi.org/10.48550/arXiv.2303.13988.
[48] Max Pellert, Clemens M. Lechner, Claudia Wagner, Beatrice Rammstedt & Markus Strohmaier, 19(5) AI Psychometrics: Assessing the Psychological Profiles of Large Language Models Through Psychometric Inventories, Perspectives on Psychological Science, 808-826 (2024).
[49] See e.g. Nicolas Yax, Hérnan Anlló & Stefano Palminteri, Studying and improving reasoning in humans and machines, 51(2) Communications Psychology 1-16 (2024).
[50] Yang Chen, Samuel N. Kirshner, Anton Ovchinnikov, Meena Andiappan & Tracy Jenkin, A Manager and an AI Walk into a Bar: Does ChatGPT Make Biased Decisions Like We Do?, Manufacturing & Service Operations Management, Articles in Advance 1-15 (2025).
[51] Yax et al., *supra* note 49.
[52] Id. („In any case the significant variability observed between models in our study indicates that a definitive answer to the question of whether LLMs exhibit cognitivie limitations and biases cannot be drawn.").
[53] See Desai & Riedl, *supra* note 5, at 29; see also John J. Horton, Large language models as simulated economic agents: What can we learn from homo silicus?, NBER Working Paper 31122 (April 2023), http://www.nber.org/papers/w31122.
[54] Thilo Hagendorff, Deception abilities emerged in large language models, Proceedings of the National Academy of Sciences 121.24 (2024): e2317967121.
[55] See generally Geraint Howells, André Janssen & Reiner Schulze (eds) Information Rights and Obligations: A Challenge for PARty Autonomy and Transactional Fairness (Ashgate 2005); see also Geraint Howells, Christian Twigg-Flesner & Thomas Wilhelmsson, Rethinking EU Consumer Law (Routledge 2018) 94-112.

Against this background, some fervent critics of the information model have claimed that mandated disclosures as a regulatory tool are broken beyond repair and should be abandoned.[58] Others have sought to save the information model by promoting "smart disclosures" building on insights from behavioral research.[59] Yet another approach would be to replace standardized information with personalized disclosures. This could make it possible to reduce the amount of information provided while at the same time increasing its relevance.[60]

The rise of agentic commerce could give an interesting new twist to this old debate. First of all, AI agents could mitigate the problem of information overload, since the capacity of AI agents to process information is not constrained by the cognitive limitations of humans. Regardless of the question of whether AI agents exhibit human-like biases in decision-making, which has yet to be conclusively answered,[61] it seems clear that AI agents can process a larger volume of information than humans. As a consequence, the use of AI agents could reduce the significance of one of the main criticisms of the information model, i.e. the problem of information overload. But AI agents cannot only process larger amounts of information; they can also filter those information items that are relevant for a particular consumption decision. Thus, in a sense, AI agents could achieve the goal that the proponents of personalized disclosures are striving for, albeit in a different way: While the idea of a personalized disclosures builds on centralized personalization by information providers, AI agents could enable decentralized personalization on the part of information recipients.

---

[56] For an overview of disclosure mandates in EU consumer law, see Christoph Busch, The Future of Pre-contractual Information Duties: From Behavioural Insights to Big Data, in Christian Twigg-Flesner (ed) Research Handbook on EU Consumer and Contract Law (Edward Elgar Publishing 2016) 221, 224-25; see also generally Peter Rott, Information Obligations and Withdrawal Rights, in Christian Twigg-Flesner (ed) The Cambridge Companion to European Union Private Law 187 (Cambridge 2010).

[57] See, e.g. Samuel Issacharoff, Martin Engel & Johanna Stark, Buttons, Boxes, Ticks, and Trust: On the Narrow Limits of Consumer Choice, in Klaus Mathis (ed) European Perspectives on Behavioural Law and Economics, Economic Analysis of Law in European Legal Scholarship 107, 118–21 (Springer 2015); Anne-Lise Sibony, Can EU Consumer Law Benefit from Behavioural Insights? An Analysis of the Unfair Practices Directive, 6 Eur Rev Private L 901, 902-03 (2014); Annette Nordhausen Scholes, Behavioural Economics and the Autonomous Consumer, 14 Camb Yearbook Eur Legal Stud 297, 306-18 (2012); Disclosure, Agents, and Consumer Protection, 167 J Institutional & Theoretical Econ 56, 61–64 (2011); see also generally Eva Maria Tscherner, Can Behavioral Research Advance Mandatory Law, Information Duties, Standard Terms and Withdrawal Rights?, 1 Austrian L J 144 (2014); Hans-W. Micklitz, Lucia A. Reisch & Kornelia Hagen, An Introduction to the Special Issue on "Behavioural Economics, Consumer Policy, and Consumer Law", 34 J Consumer Pol 271 (2011).

[58] See, e.g. Omri Ben-Shahar & Carl E. Schneider, More Than You Wanted to Know (Princeton University Press 2014) 183; see also Omri Ben-Shahar & Carl E. Schneider, Coping with the failure of mandated disclosure 11 Jerusalem Review of Legal Studies 83-95 (2015).

[59] See, e.g. Oren Bar-Gill, Defending (smart) disclosure: A comment on More than You Wanted to Know 11 Jerusalem Review of Legal Studies 75 (2015).

[60] Ariel Porat and Lior Strahilevitz, Personalizing default rules and disclosure with big data, 112 Michigan Law Review 1417 (2014); Christoph Busch, Implementing Personalized Law: Personalized Disclosures in Consumer Law and Data Privacy Law, 86 U. Chi. L. Rev. 309 (2019).

[61] See *supra* B.I.

What follows from this for the future of pre-contractual information requirements in consumer law? Two different policy responses could be envisaged. One option would be to reduce the number of mandated disclosures for transactions involving AI agents. In justification of such an approach, one could argue that AI agents are able to obtain relevant information from many different sources (e.g. customer reviews, user forums, product tests, online media). From this perspective, "algorithmic consumers" would need less protection than human consumers. This approach, however, would involve a significant shift of risk to the detriment of the consumer. Whether the consumer is actually able to take an informed decision would in such a scenario very much depend on the capabilities of the specific AI agent used by the consumer.

Therefore, a different reaction of consumer law to the rise of AI agents might be preferrable. Instead of reducing the number of pre-contractual disclosures, traders could also be required to provide AI agents with more detailed information. For human consumers, this would be counterproductive as it would lead to information overload. Providing more information to human consumers would not improve the decision quality, but possibly worsen it. In contrast, AI agents will be able to process the detailed information and filter out the data that is relevant to the consumer's pre-set preferences.

It appears that the EU has recently begun to adapt the information paradigm to the "machine age". Some of the latest EU legal acts stipulate that important information must be provided in an "easily accessible and machine-readable format".[62] Another example of information that seems to be aimed at AI agents rather than humans is the European digital product passport (DPP), as envisaged by the EU Ecodesign for Sustainable Products Regulation.[63] The DPP is essentially a digital identity card for products, a digital file that contains a structured set of product-related data which can be accessed electronically, e.g. via a QR code.[64] The DPP, which will become mandatory for various categories of products between 2026 and 2030, will include information about materials used, manufacturing methods, sustainability features, and guidelines for repair and recycling. For human consumers, such information is probably too complex. In contrast, AI agents could easily process this information and help consumers in making more informed decisions related to sustainable and circular consumption.

### III. AI Agents as Choice Engines

Another area of consumer law that may have to be reconsidered in the light of agentic commerce, are the rules aimed at protecting consumers against manipulation by unfair commercial practices. These include not only the prohibitions of misleading and aggressive practices enshrined in the Unfair Commercial Practices Directive (UCPD), but also more recent regulations against dark patterns, such as Art. 25 Digital Services Act (DSA). These regulations have in common that they are intended to protect consumers from manipulation and the exploitation of human weaknesses. Therefore, the design of these rules increasingly takes into account insights from cognitive psychology about heuristics and biases that distort human decision-making.

---

[62] See e.g. Art. 14(1) Digital Services Act. Another reason for the requirement to provide the information in a machine-readable format could be to enable automated monitoring by enforcement authorities.
[63] Regulation (EU) 2024/1781 of the European Parliament and of the Council of 13 June 2024 establishing a framework for the setting of ecodesign requirements for sustainable products, amending Directive (EU) 2020/1828 and Regulation (EU) 2023/1542 and repealing Directive 2009/125/EC, OJ L, 2024/1781, 28.6.2024.
[64] See, e.g. Charlotte Ducuing & René Herbert Reich, Data governance: Digital product passports as a case study, 24(1) Competition and Regulation in Network Industries, 3, 7-8 (2023).

In the Custobot Economy, these rules may have to be recalibrated as the vulnerability of AI agents to misleading and aggressive commercial practices may differ from that of humans.[65] Although more empirical research is needed here, it can be assumed that AI agents are less vulnerable to at least some of the dark patterns that specifically target human behavioral vulnerabilities. Presumably, AI agents will not be fooled by the color or size of buttons on a website. It may also be assumed that AI agents are less receptive to emotional advertising messages.

More generally, AI agents could help consumers make better decisions and overcome behavioral biases. The question of whether and how consumers should be "nudged" towards better decisions through appropriate choice architecture has been the subject of broad discussion.[66] The advent of AI agents could take the idea of nudging to the next level. AI agents could function as Choice Engines that help consumers to overcome both informational deficits and behavioural biases.[67] In steering consumers' decision-making, Choice Engines may exhibit different levels of paternalism. For example, if a consumer repeatedly orders unhealthy food, a moderately paternalistic AI agent could suggest healthier alternatives. Because AI agents collect a large amount of data about the preferences and vulnerabilities of individual consumers, they can also deliver highly personalized nudges. This could solve the problem that standardized nudges often lack precision.[68]

A more provocative idea is that Choice Engines might also take account of externalities in their decision-making.[69] In doing so, they might nudge users to make choices that reduce negative effects for others, e.g. by indicating the respective carbon footprint of different options. An even more radical idea would be for an AI agent to refuse a transaction if it exceeds a certain level of greenhouse gas emissions. This example underscores that AI agents could not only function as a tool of consumer empowerment, but also as a highly paternalistic instrument for restricting freedom of choice.

**IV. Digital Vulnerability 2.0**

The rise of agentic commerce could also give a new twist to the policy debate about digital vulnerability. According to a growing strand of literature, consumer vulnerability in digital environments is a structural or systemic phenomenon resulting in particular from the design of digital choice architectures.[70] The use of AI agents could reduce vulnerability to digital choice architectures that are designed to exploit human weaknesses. However, this does not mean that AI agents are invulnerable against manipulation. Instead of manipulating human consumers, malevolent sellers might seek to manipulate AI agents by exploiting technical vulnerabilities.[71] While AI agents may be immune to some weaknesses that human consumers exhibit, there could be new AI-specific vulnerabilities that do not occur in the context of human decision making.

---

[65] See *supra* C.I.
[66] See, e.g. Cass R. Sunstein & Richard H. Thaler, Libertarian Paternalism is Not an Oxymoron 70 U Chi L Rev 1159 (2003); Richard H. Thaler & Cass R. Sunstein, Nudge. Improving Decisions about ealth, Welath, and Happiness (Yale UP 2008); Alberto Alemanno & Anne-Lise Sibony (Eds) Nudge and the Law: A European Perspective (Hart 2015); Daniel M. Hausman & Brynn Welch, Debate: To Nudge or Not to Nudge, 18 The Journal of Political Philosophy 123 (2010).
[67] Cass R. Sunstein, Choice engines and paternalistic AI, 11 Humanit Soc Sci Commun 888 (2024); see also Cass R. Sunstein, Brave New World? Human Welfare and Paternalistic AI (July 29, 2024), Theoretical Inquiries in Law (forthcoming), https://ssrn.com/abstract=4908836.
[68] Stuart Mills, Personalized nudging, 6(1) Behavioural Public Policy, 150–159 (2022).
[69] See Cass R. Sunstein, *supra* note 67.
[70] For an overview, see Camilla Crea and Alberto De Franceschi (eds), The New Shapes of Digital Vulnerability in European Private Law (Nomos 2024); see also Natali Helberger, Betül Kas, Hans-W. Micklitz, Monika Namysłowska, Laurens Naudts, Peter Rott, Marijn Sax, Michael Veale, Digital Fairness for Consumers, BEUC Report (May 2024), p. 12 (arguing that the concept of digital vulnerability is characterised by three aspects, "its relational nature, its architectural nature, and the erosion of privacy"). Other authors refer to "systemic vulnerability", see e.g. Christine Riefa, Protecting vulnerable consumers in the digital single market, (2022) European Bustiness Law Review, 33, 607-634 (arguing that consumer vulnerability is created by the system design, in particular the design of online choice architectures).

In particular, AI agents could be vulnerable to cybersecurity exploits and adversarial attacks that trick Custobots into a transaction. The risks associated with adversarial attacks are well known from visual classification algorithms which are used, for example, by self-driving cars in order to "see" traffic signs. Research in this field has shown that it is possible to confuse computer vision systems by slight alterations to an image invisible to humans ("adversarial image perturbations") or by manipulating physical objects. For example, by attaching some stickers to a stop sign a deep neural network-based classifier can be fooled into "thinking" that it was "looking" at a speed limit sign.[72]

This example underlines that AI systems "see" the world differently from humans and can therefore be manipulated in surprising ways. In the context of agentic commerce, one could imagine adversarial images being used in online shops to manipulate Custobots. For this purpose, subtle image-specific modulations of individual pixels that are designed to alter image categorization could be used. While such "adversarial images" would have no impact on human consumers, they could act as a new form of AI-specific dark pattern for manipulating Custobots. In addition, recent research on AI agents underlines the risk of prompt injection atttacks, which embed malicious commands to alter the behaviour of AI agents.[73] For example, AI agents could be hijacked to leak confidential user data or trigger harmful actions (e.g. unauthorized transfers of funds).[74]

## D. Towards a Consumer Law that Works for Humans and Machines

Based on the preceding analysis, this Part briefly outlines some initial considerations for how a future consumer law could look like that works for both humans and machines. As the analysis in the previous Parts has shown, the rise of AI agents requires a number of adjustments to the current legal and technical infrastructure of electronic commerce.

Work is already underway to make the existing legal framework fit-for-purpose for the coming wave of agentic commerce. In July 2024, the United Nations Commission on International Trade Law (UNCITRAL) adopted a *Model Law on Automated Contracting* (MLAC) which provides a legal framework for the use of automated systems in international contracts.[75]

---

[71] Van Loo, *supra* note 6, 841 (arguing that sellers might "adopt misperception tactics against AIs similar to those they deploy for consumers").
[72] Kevin Eykholt, Ivan Evtimov, Earlence Fernandes, Bo Li, Amir Rahmati, Chaowei Xiao, Atul Prakash, Tadoyoshi Kohno & Dawn Song, Robust Physical-World Attacks on Deep Learning Models (Jul. 27, 2017), https://doi.org/10.48550/arXiv.1707.08945; see also Vijay Veerabadran, Josh Goldman, Shreya Shankar, Brian Cheung, Nicolas Papernot, Alexey Kurakin, Ian Goodfellow, Jonathon Shlens, Jascha Sohl-Dickstein, Michael Mozer & Gamaleldin F. Elsayed, Subtle adversarial image manipulations influence both human and machine perception, Nat. Commun. 14, 4933 (2023), https://doi.org/10.1038/s41467-023-40499-0.
[73] Peter Yong Zhong, Siyuan Chen, Ruiqi Wang, McKenna McLall, Ben L. Titzer, Heather Miller and Phillip B. Gibbons, RTBAS: Defending LLM Agents Against Prompt Injection and Privacy Leakage (Feb. 14, 2025), https://doi.org/10.48550/arXiv.2502.08966.
[74] Id.
[75] United Nations Commission on International Trade Law, Model Law on Automated Contracting (July 2024), https://uncitral.un.org/sites/uncitral.un.org/files/mlac_en.pdf.

However, the MLAC is primarily designed to apply to commercial contracts (B2B) rather than consumer contracts. The same applies to the *Principles for AI in Contracting* (PAIC), an academic proposal for a set of legal principles for automated contracting.[76] In contrast, the *European Law Institute Guiding Principles and Model Rules on Digital Assistants for Consumer Contracts* (ELI DACC Model Rules) focus specifically on the use of AI agents for consumer contracts.[77]

The following sections do not aim to provide an in-depth comparative analysis of the different regulatory regimes. Instead, they seek to conclude this Article with a brief look at three overarching questions: How can consumer law enable and facilitate the use of AI agents? How should AI agents be protected against manipulation? And last but not least: How should consumers be protected against manipulation by AI agents?

**I. Enabling the Custobot Economy**

In view of the numerous advantages that AI agents offer for consumers and the potential for consumer empowerment, the first question to be addressed is how consumer law can help to enable the use of AI agents. This consideration is based on the assumption that an innovation-friendly consumer law has a dual function: On the one hand, the regulatory design of consumer law should enable the use of AI agents to empower consumers (enabling function). On the other hand, consumer law should ensure that consumers are adequately protected when AI agents are used (protective function).[78] The legislator's task is to strike a balance between these two objectives.

First of all, the legal recognition of contracts concluded via AI agents should be clarified.[79] Twenty-five years ago, in the early days of the commercial Internet, Article 9 of the E-Commerce Directive[80] laid the contractual foundations for electronic commerce by stipulating that Members States should allow contracts to be concluded by electronic means. Today, as we are standing at the threshold of agentic commerce, EU law should move to the next level by clarifying that all national legal systems should allow contracts to be concluded by automated means via AI agents.

But this is not enough. EU consumer law should also clarify that consumers have the right to use AI systems for concluding contracts with traders.[81] Any contractual terms or agreements which restrict the use of AI agents should not be binding on the consumer.[82] In a similar way, traders should not create any technical obstacles that prevent the use of AI agents.[83] In other words, traders should be required to use an AI-friendly design for their websites or apps.[84] The ELI DACC Model Rules refer to this as the "no barrier" principle.[85]

---

[76] Christiane Wendehorst, Discussion Draft: Principles for AI in Contracting (Version 2.1), 13 Journal of European Consumer and Market Law 43-60 (2024).

[77] European Law Institute, Guiding Principles and Model Rules on Digital Assistants for Consumer Contracts (forthcoming March 2025).

[78] Cf. European Law Institute, EU Consumer Law and Automated Decision-Making (ADM): Is EU Consumer Law Ready for ADM?, Interim Report of the European Law Institute (Dec. 18, 2023) https://www.europeanlawinstitute.eu/fileadmin/user_upload/p_eli/Publications/ELI_Interim_Report_on_EU_Consumer_Law_and_Automated_Decision-Making.pdf.

[79] See Art. 17 ELI DACC Model Rules („An algorithmic contract shall not be denied validity or enforceability solely because a digital assistant was used, irrespective of whether only one party or both parties used digital assistants."); see also Art. 3 PAIC („A contract should not be denied validity or enforceability on the sole ground that it has been concluded through electronic agents, irrespective of the extent to which the output is controlled by the operator of the electronic agent or of the general degree of predictability of the output.").

[80] Directive 2000/31/EC of the European Parliament and of the Council of 8 June 2000 on certain legal aspects of information society services, in particular electronic commerce, in the Internal Market, OJ L 178, 17.7.2000, p. 1-16.

[81] See Art. 3(1) ELI DACC Model Rules („Consumers have the right to use a digital assistant for their contractual relations with a business.").

Certainly, this principle does not apply without exceptions. In certain circumstances, a trader may have a legitimate interest in preventing the use of machine customers. Think, for example, of AI agents being used for the automated purchasing of tickets for sports events or concerts in order to circumvent any limit on the number of tickets that a person can buy or any other rules applicable to the purchase of tickets.[86] With the widespread use of AI agents such scenarios will occur more frequently and courts will have to draw the line between acceptable and unacceptable use of AI agents.[87]

## II. Protecting Custobots Against Manipulation

As discussed above, malevolent sellers might seek to manipulate Custobots by exploiting technical vulnerabilities in order to trick the AI agent into a transaction.[88] So far, EU consumer law is lacking any specific rules that protect AI agents from manipulation. Art. 23(a) ELI DACC Model Rules contains an innovative proposal for such a rule: „A business must not use the structure, design, function, or manner of operation of their online interface in a way that is likely to materially distort or impair the ability of a digital assistant to perform its functions." It is obvious that the wording of the proposal is modelled on Art. 25(1) of the Digital Services Act, which prohibits providers of online platforms from using dark patterns. The analogy seems fitting. Just as human consumers can be manipulated by dark patterns, AI agents can be manipulated by the "structure, design, function, or manner of operation" of online interfaces.

However, the ELI proposal deviates from its model in one important respect. Art. 25(1) DSA speaks of dark patterns impairing the ability "to make free and informed decisions". This formulation is not used in the ELI proposal. It would probably have been too bold to claim that AI agents make "free decisions", as this would imply that AI agents have something like a "free will". Instead, the ELI proposal only speaks of an impairment of the AI agent's ability „to perform its functions". This formulation adequately reflects the instrumental character of AI agents.

The adoption of a provision drafted along the lines of Art. 23(a) ELI DACC Model Rules – possibly as part of the forthcoming EU Digital Fairness Act[89] – is only the first step. To ensure effective consumer protection, we need to deepen our understanding of the vulnerabilities of AI agents. While research on dark patterns has made significant progress in recent years,[90] research on the behavior of AI agents and on "machine psychology" is still in the early stages. Further research is needed, for example, to develop taxonomies of malicious design patterns that are likely to manipulate AI agents.

---

[82] Cf. Art. 3(2) ELI DACC Model Rules.
[83] Cf. Art. 3(3) ELI DACC Model Rules.
[84] Art. 3(5) ELI DACC Model Rules provides for an exception for small and micro-enterprises.
[85] ELI DACC Model Rules, Introduction.
[86] See Unfair Commercial Practices Directive, Annex I, No. 23a.
[87] An early example of a similar debate is the controversy about sniper bots on eBay, see Matthew Backus, Tom Blake, Dimitriy V. Masterov & Steven Tadelis, Is Sniping A Problem For Online Auction Markets?, NBER Working Paper 20942, February 2015, https://www.nber.org/papers/w20942; see also Stefan Leible & Olaf Sosnitza, Sniper-Software und Wettbewerbsrecht. Zur Vertrags- und Lauterkeitsrechtlichen Beurteilung automatisierter Gebote bei Internet-Auktionen, 19 Computer und Recht 344 (2003).
[88] See *supra*, C.IV.
[89] Cf. Christoph Busch & Amelia Fletcher, Shaping the Future of European Consumer Protection: Towards a Digital Fairness Act?, CERRE Issue Paper (December 2024), https://cerre.eu/publications/shaping-the-future-of-european-consumer-protection-towards-a-digital-fairness-act/.

**III. Protecting Humans Against Custobots**

While the focus of this article is on how consumer law should be adapted to the strengths and weaknesses of the new species of "algorithmic consumers", the analysis would be incomplete if it did not at least briefly consider the risks for human consumers when using AI agents. This issue, which has already been addressed by several scholars,[91] could even become a key policy theme in the near future.[92] In this respect, too, important structural changes in consumer law are to be expected. More specifically, the rise of AI agents may necessitate a shift of focus in consumer law in order to ensure effective protection of human consumers.

Currently, EU consumer law puts its focus around the "magical moment" when the contract is concluded.[93] This is the time – shortly before and after the conclusion of a consumer contract – when the full arsenal of consumer protection instruments comes into play: pre-contractual information duties, documentation requirements, form requirements for the conclusion of consumer contracts, and withdrawal rights. Together, these instruments aim to ensure a free and informed decision.

However, in the Custobot Economy, there may be no human consumer involved at this point, but only an AI agent acting on behalf of the consumer. The human decision-making process takes place at an earlier stage, when the consumer deploys the AI agent, defines parameters, and sets preferences for the operation of the AI agent (e.g. defining a maximum value for transactions, or specifiying particular suppliers).[94] Therefore consumer protection has to step in at an earlier stage and it must address the design requirements for AI agents.[95] As a consequence, we might see a shift of focus away from consumer contract law towards a design-based approach to consumer protection. This trend is also reflected in the work of the European Law Institute on AI agents. The ELI DACC Model Rules combine design requirements for AI agents with contract law rules. The idea here is that „the objective of ensuring high levels of consumer protection in respect of digital assistants is best achieved through a combination of *ex ante* requirements and *ex post* rights."[96]

---

[90] See e.g. Jamie Luguri & Lior Jacob Strahilevitz, Shining a Light on Dark Patterns, 13(1) Journal of Legal Analysis, 43-109 (2021); Colin M. Gray, Natalia Bielova, Cristiana Santos & Thomas Mildner, An Ontology of Dark Patterns: Foundations, Definitions, and a Structure for Transdisciplinary Action (Sep. 18, 2023), https://doi.org/10.48550/arXiv.2309.09640; Martin Brenncke, Regulating Dark Patterns, 14(1) Notre Dame Journal of International & Comparative Law 39-79 (2024).
[91] See e.g. Kolt, *supra* note 4; Desai & Riedl, *supra* note 5.
[92] Cf. Zittrain, *supra* note 10 (warning of „potentially devastating consequences" if no effective measures are introduced to contral AI agents now).
[93] See Hans Schulte-Nölke, EC Law on the Formation of Contract – from the Common Frame of Reference to the 'Blue Button', 3(3) European Review of Contract Law, 332-349 (2007) (arguing that EU law exhibits a high level of regulatory density before and after the conclusion of contract while the moment of conclusion, the "magical moment" when the contract becomes binding, is almost not regulated at all).
[94] See Christian Twigg-Flesner, Consumers and Digital Delegates, in Larry A. DiMatteo, Cristina Poncibó & Geraint Howells (eds), The Cambridge Handbook of AI and Consumer Law: Comparative Perspectives (Cambridge University Press 2024) 75, 86.
[95] See Arts. 4-10 ELI DACC Model Rules (setting out design requirements for AI agents).
[96] ELI DACC Model Rules, Introduction.

# E. Conclusion

Scholars have been arguing already for some time that the one-size-fits all model of consumer law no longer corresponds to the complexities of modern consumer markets and that a greater differentiation between different types of consumers is required.[97] The rise of AI agents brings a new twist to this argument. With the rise of agentic commerce, the complexity of the consumer market is increasing even further. In the new Custobot Economy, human consumers will co-exist with AI-driven "algorithmic consumers". This will affect almost all aspects of consumer law, from the average consumer concept to the information paradigm and the concept of digital vulnerability. Adapting consumer law rules to the coming wave of Custobots will require a nuanced understanding of the strengths and weaknesses of AI agents. Policymakers should not wait for the next "iPhone moment"[98] to address these challenges; they should start now creating a consumer law that works for both humans and machines.

---

[97] Vanessa Mak, How Can Consumer Interest be Protected When Consumer Identities are Increasingly Diffuse?, in Hans-W. Micklitz & Christian Twigg-Flesner (eds) The Transformations of Consumer Law and Policy in Europe (Hart 2023), 43; see also Vanessa Mak, Legal Pluralism in European Contract Law (Oxford UP 2020) 119.
[98] Cf. Bernard Marr, Is this AI's iPhone moment?, Forbes (Jul. 8, 2024), https://www.forbes.com/sites/bernardmarr/2024/07/08/is-this-ais-iphone-moment/